\documentstyle[12pt]{article}
\newcommand{\be}{\begin{equation}}
\newcommand{\ee}{\end{equation}}
\newcommand{\bea}{\begin{eqnarray}}
\newcommand{\eea}{\end{eqnarray}}
\newcommand{\ba}{\begin{array}}
\newcommand{\ea}{\end{array}}

\begin{document}
\baselineskip = 18 pt
 \thispagestyle{empty}
 \title{
 \vspace*{-2.5cm}
 \begin{flushright}
 \begin{tabular}{c c c c}
 \vspace{-0.3cm}
 & {\normalsize CERN-TH.7321/94}\\
 \end{tabular}
 \end{flushright}
 \vspace*{0.4cm}
Soft Supersymmetry Breaking
Parameters and Minimal SO(10) Unification$^*$
\\ }
 \author{   M. Carena and  C.E.M. Wagner\\
{}~\\
{}~\\
CERN Theory Division,
1211 Geneva 23, Switzerland\\
{}~\\
{}~\\
{}~\\
 }
\date{
\begin{abstract}
The minimal supersymmetric SO(10) model, in which not only
the gauge but also the third generation fermion Yukawa couplings
are unified, provides a simple and highly predictive theoretical
scenario for the understanding of the origin of the low energy
gauge interactions and fermion masses. In the framework of the
Minimal Supersymmetric Standard Model with universal soft
supersymmetry breaking parameters at the grand unification scale,
large values of the universal gaugino mass $M_{1/2} \geq 300$ GeV
are needed in order to induce a proper breakdown of the electroweak
symmetry. In addition, in order to obtain
acceptable experimental values
for both the pole bottom mass and the $b \rightarrow s \gamma$ decay
rate, even larger values of the gaugino masses are required.
The model is strongly constrained by theoretical and phenomenological
requirements and a heavy top quark, with mass $M_t \geq 170$ GeV,
is hard to accomodate within this scheme. We show, however,
 that it is sufficient
to  relax the condition of universality of the scalar soft supersymmetry
breaking parameters at the grand unification scale
to be able to accommodate
a top quark mass $M_t \simeq 180$ GeV. Still, the
requirement of a heavy top
quark demands  a very heavy squark spectrum,
unless specific relations between the soft supersymmetry breaking
parameters are fulfilled.\\
 \begin{flushleft}
 {\normalsize CERN-TH.7321/94} \\
 {\normalsize June 1994}\\
{}~\\
$*$
 Talk
 presented by C.E.M.$\;$ Wagner$\;$
 at the 2nd $\;$ IFT $\;$Workshop $\;$on $\;$Yukawa Couplings
 and the Origins of Mass, Gainesville, Florida, Feb. 1994. To appear in
 the Proceedings.
\end{flushleft}
\end{abstract}}
\maketitle
\newpage
\section{Introduction}

The Standard Model provides a very good understanding of the
strong and electroweak interactions and, so far, it has withstood all
the experimental onslaughts. Yet, it leaves
a host of open questions, which  require, to be answered,
the presence of
an underlying structure with a larger symmetry content than
the one sufficient to describe the physics
 at present accelerator energies.
In particular, an explanation to the  origin of
forces and fermion masses, as well as the
hierarchy  between the Planck and the weak scale is still lacking.
 Supersymmetric theories have the potential of dealing with these
problems and, at the same time, of remaining compatible with the low
energy data. This is
why, in spite of the lack of
experimental evidence for supersymmetry, most
theories beyond the Standard Model include this symmetry
at some stage. In the simplest supersymmetric theories,
gauge invariant, soft supersymmetry breaking terms are present,
which yield masses to the unobserved supersymmetric particles,
while the standard model fermion and boson fields acquire masses
through the usual Higgs mechanism \cite{DGR}.
In addition, supersymmetry
ensures the stability of the hierarchy between the weak and
the Planck scales.  Moreover, in the minimal
supersymmetric models the weak scale and the scale of
supersymmetry breaking are interrelated, and a natural
explanation of the scale of electroweak symmetry breaking
may only be achieved if the supersymmetric particle masses
are smaller than,  or of the order of, 1 TeV.

It has  recently been realized that the weak and
strong gauge couplings determined by the most recent
measurements at the LEP experiments are consistent with the
 unification of
gauge couplings within the minimal supersymmetric
standard model \cite{unif}.
The presence of gauge coupling unification has strongly revived
the interest in softly broken supersymmetric theories,
in particular, in the Minimal Supersymmetric extension of
the Standard Model.
Much work has been done in understanding
the influence of threshold corrections at low and high energy
scales and the impact of the precise supersymmetric spectrum on the
predictions for the
strong gauge coupling \cite{LP}. In particular, it has been
shown that the low energy threshold corrections are strongly
dependent on the Higgsino and gaugino masses, and weakly
dependent on the squark and sfermion masses \cite{CPW}. Moreover, the
relevance of the experimental correlation between the top
quark mass and the weak mixing angle in  the obtention of
the strong gauge
coupling predictions has been emphasized. As we shall discuss
below, due to this correlation, a heavy top quark,
with mass $M_t > 150 \;(170)$ GeV,
is associated with values of the strong gauge coupling
$\alpha_3(M_Z) \geq 0.114 \;(0.117)$. The lower bound on
$\alpha_3(M_Z)$ increases
for larger values of the top quark mass.

Furthermore, the question of fermion masses
may also find a natural explanation within
minimal supersymmetric Grand Unified Theories
(see, for example, Refs. \cite{Hall1},\cite{fermion}).
In fact, minimal supersymmetric GUTs give a natural
explanation for the heaviness of the top quark: The
condition of bottom--tau Yukawa coupling unification
\cite{Ramond},\cite{KLN} requires
large values of the top quark Yukawa coupling, $h_t$,
at the grand unification scale, $M_{GUT}$,  in order to contravene
the strong gauge coupling effects on the running of the bottom Yukawa
coupling \cite{LP}--\cite{Hall1},\cite{BABE}.
Moreover,  if the top quark Yukawa coupling
acquires large values at the grand unification scale,
$Y_t = h_t^2/4\pi \geq 0.1$, its low energy value
is completely determined by the infrared fixed point
structure of the theory \cite{IR}--\cite{Dyn}.
Indeed, for small and moderate values of $\tan \beta$
--the ratio of vacuum expectation
values of the Higgs fields-- for which the effects of the bottom
quark Yukawa coupling in the running of $h_t$ may be safely neglected,
 the infrared quasi-fixed
point value of the running top quark mass in the $\bar{MS}$
scheme is approximately  given by

\begin{equation}
m_t (M_t)^{IR}  \simeq \; 196 \;
 {\mathrm{ GeV} } \left[1 + 2 \; (\alpha_3(M_Z) - 0.12)
\right] \sin\beta
\label{eq:fixmt}
\end{equation}
where the pole mass is related to the running mass by \cite{Ramond2}
\begin{equation}
M_t \simeq m_t(M_t) \left[1 + 4\alpha_3(M_t)/3\pi +
11 (\alpha_3(M_t)/\pi)^2\right].
\label{eq:runpomt}
\end{equation}
A careful analysis shows that,
for small and moderate values of $\tan \beta$
   the conditions of gauge
and bottom--tau Yukawa coupling unification lead to a top quark
mass which, for the currently acceptable values for the
bottom mass and the electroweak parameters,
 is within $10 \%$ of its infrared fixed point
value \cite{LP2},\cite{BCPW}.

In general,
the condition of bottom--tau Yukawa coupling unification
determines the value of the top quark Yukawa coupling and, in the small
and moderate $\tan \beta$  regime, implies
a strong  attraction of the top quark
mass to its infrared fixed point. Observe that the infrared fixed
point solution  does not provide a direct prediction for the
top quark mass, but only a strong correlation between $M_t$
and $\tan\beta$. A prediction for both
the top quark mass and $\tan\beta$ may only be obtained in theories
with a richer symmetry structure than the one provided by
the minimal supersymmetric  SU$(5)$ theory.
 In this respect,
the minimal SO(10) model, where all Yukawa couplings proceed from
a common coupling of the 16  representation of matter fields
with a 10 representation of Higgs fields, provides a natural
extension of the SU(5) scenario  \cite{fermion},\cite{OP2}--\cite{OP}.
Since the top and bottom Yukawa couplings are of the same order,
the hierarchy of top and bottom masses is due to a large value
of the ratio of vacuum expectation values:

\begin{equation}
\tan\beta \simeq \frac{m_t(m_t)}{m_b(m_t)} \simeq {\cal{O}}(50).
\label{eq:lartanb}
\end{equation}
The condition of bottom--tau Yukawa unification is implicit
within this scheme, therefore,
the infrared fixed point attraction is potentially present
and the top quark mass tends to get larger values. However, since
 the bottom and the top Yukawa couplings are of the
same order, the bottom Yukawa effect is  sufficiently
strong by itself to partially contravene the strong gauge
coupling effects on its renormalization group running.
Then, the top quark Yukawa coupling at the grand
unification scale tends to be smaller than for moderate values
of $\tan\beta$ and
the infrared fixed point attraction becomes
weaker. The top quark mass
prediction becomes much more sensitive to the
actual value of the bottom mass and the strong gauge coupling.
Hence, the strong predictivity expected from the combination of
large values of $\tan\beta$ and the infrared fixed point attraction
is not actually realized within this scheme.

Moreover, for the large values of $\tan\beta$ implied by the
above condition, Eq. (\ref{eq:lartanb}), large corrections
to the running bottom mass induced by the supersymmetry
breaking sector of the theory are present
\cite{fermion},\cite{Banks}--\cite{Ralf}.
These corrections,
which may be as large as $50 \%$ of the
bottom mass value, make the top quark mass predictions highly dependent
on the nature of the supersymmetry breaking sector of the
theory. For instance, if the soft supersymmetry breaking parameters
are such that these corrections are negligible,
larger values of the top quark mass $M_t \geq 170$ GeV are
preferred. In the case of universal soft supersymmetry
breaking parameters at the grand unification scale,
these corrections are, instead,
large, and lower values of the top quark
mass, $M_t \leq 170$ GeV, are preferred \cite{wefour}.
Hence, no prediction for
the top quark mass may be  obtained unless a specific
framework for the breakdown of supersymmetry is given.

In this talk, we  first concentrate on the simplest
supersymmetry breaking scenario, with universal soft supersymmetry
breaking parameters at the grand unification scale, and we
describe the
phenomenological and theoretical constraints arising in this model
(for a detailed discussion of similar issues at low values of
$\tan\beta$, we refer the reader to
 \cite{COPW} and \cite{wetwo}).  We shall show
that strong correlations between the different sparticle masses appear
within this scheme. Furthermore, a lower bound on the squark
and gaugino masses is obtained from the requirement of a proper
SU(2)$_L$ $\times$ U(1)$_Y$ breakdown.
We shall also
discuss the corrections to the bottom mass, its correlation with
the supersymmetric contributions to the $b \rightarrow s \gamma$
decay rate and its implication for the top quark mass predictions.
Finally, we shall briefly describe the implications of relaxing
the condition
of universality of the soft supersymmetry
breaking parameters, both
in the Higgs and the sfermion sector of the theory.

\section{Gauge Coupling Unification}

In order to analyse the condition of gauge coupling unification,
the experimental correlation between the top quark pole mass and
the weak mixing angle  should be
considered. Indeed, taking as input values the Fermi constant,
the value of the $Z$--boson mass  $M_Z$, and the value of
$\alpha_{em}(M_Z)$, in the modified $\bar{MS}$ scheme
a correlation between the top quark mass and
$\sin^2\theta_W(M_Z)$ is induced through the top quark mass
dependent radiative corrections to the weak mixing angle \cite{LP},

\begin{equation}
\sin^2\theta_W(M_Z) = 0.2324 - 10^{-7} \left( M_t^2 -
(138 \;\;{\mathrm{ GeV }})^2 \right) {\mathrm{ GeV }}^{-2} \pm 0.003\; .
\label{eq:sint}
\end{equation}

In addition, in order to fully understand the implications
of gauge coupling unification, a few words about the
supersymmetric threshold corrections to the gauge couplings
should be said.
For a given supersymmetric spectrum, and a fixed
value of the weak mixing angle,  the value of $\alpha_3(M_Z)$,
determined by the gauge coupling unification condition,
is given by
\begin{eqnarray}
\frac{1}{\alpha_3(M_Z)}  &=&
\frac{ \left(
b_1 - b_3 \right)}
{ \left(
b_1 - b_2 \right)} \left[
\frac{1}{\alpha_2(M_Z)}
 + \gamma_2  + \Delta_2 \right]
-
\frac{ \left(
b_2 - b_3 \right)}
{ \left(
b_1 - b_2 \right)}  \left[
\frac{1}{\alpha_1(M_Z)}
 + \gamma_1  + \Delta_1 \right]
\nonumber\\
& - & \gamma_3  - \Delta_3
+ \Delta^{Sthr}\left(\frac{1}{\alpha_3(M_Z)}\right) ,
\label{eq:pred3}
\end{eqnarray}
where
\begin{equation}
\Delta^{Sthr}\left(\frac{1}{\alpha_3(M_Z)}\right)
= \frac{19}{28 \pi}
\ln \left( \frac{T_{SUSY}}{M_Z}\right)
\label{eq:Tsusy}
\end{equation}
is the contribution
to $1/\alpha_3(M_Z)$ due to the inclusion of the supersymmetric
threshold corrections at the one--loop level, $\gamma_i$
includes the two--loop corrections to the value of $1/\alpha_i(M_Z)$,
$\Delta_i$ are correction constants that allow a transformation of  the
gauge couplings from the minimal $\bar{MS}$ scheme to the
dimensional reduction scheme $\bar{DR}$, more appropriate for
supersymmetric theories, and $b_i$ are the supersymmetric beta
function coefficients associated to the gauge coupling $\alpha_i$.
As becomes clear from Eq. (\ref{eq:Tsusy}), the effective threshold
scale $T_{SUSY}$ gives a parametrization of the size of the
supersymmetric threshold corrections to the gauge couplings and
it would coincide with the overall mass scale $M_{SUSY}$ only if
all supersymmetric particles were degenerate in mass.

In order to study the dependence of $T_{SUSY}$ on the
different sparticle masses,
we define
$m_{\tilde{q}}$,
$m_{\tilde{g}}$,
$m_{\tilde{l}}$,
$m_{\tilde{W}}$,
$m_{\tilde{H}}$ and $m_H$
as the characteristic masses of the
squarks, gluinos, sleptons, electroweak gauginos,
Higgsinos and  the heavy Higgs doublet,
respectively. Assuming different values for all these
mass scales, we derive an expression for
the effective supersymmetric threshold $T_{SUSY}$, which is given
by \cite{CPW}
\begin{equation}
T_{SUSY} =
m_{\tilde{H}}
\left( \frac{m_{\tilde{W}}
}{m_{\tilde{g}}}
\right)^{28/19}
\left[
\left( \frac{m_{\tilde{l}}}{m_{\tilde{q}}}
\right)^{3/19}
\left( \frac{m_H}{m_{\tilde{H}}}
\right)^{3/19}
\left( \frac{m_{\tilde{W}}}{m_{\tilde{H}}}
\right)^{4/19} \right] .
\label{eq:susym}
\end{equation}
The above relation holds  whenever
 all the particles involved have  masses
 $m_{\eta} > M_Z$. If, instead,
 any of the sparticles or the heavy Higgs boson
has a mass $m_{\eta} < M_Z$, it  should be replaced
by $M_Z$ for the purpose of computing the
supersymmetric threshold corrections to $1/ \alpha_3(M_Z)$.
{}From Eq. (\ref{eq:susym}), it follows that
$T_{SUSY}$ has only a slight dependence on the
squark, slepton and heavy Higgs masses and a very strong dependence on
the overall Higgsino mass, as well as on the ratio of
masses of the gauginos associated with the
electroweak and strong interactions.
In Table 1, we show the predictions for the strong gauge
coupling, for different values of $\sin^2\theta_W(M_Z)$ and
the supersymmetric threshold scale, together with the approximate
value of the top quark pole mass associated with each value of
 $\sin^2\theta_W(M_Z)$.
\\
\baselineskip=18pt
\begin{center}
\begin{tabular}{|c|c|c|c|}
\hline \  $M_t$ [GeV] &$\sin^2\theta_W(M_Z)$
&$\alpha_3(M_Z)$ for $T_{SUSY} = 1$ TeV
 &$\alpha_3(M_Z)$ for $T_{SUSY} = 100$ GeV
\\ \hline
140  &$0.2324$
 &0.116  &0.123
\\ \hline
170  &$0.2315$
 &0.119  &0.127
\\ \hline
195  &$0.2305$
 &0.123  &0.131
\\ \hline
\end{tabular}
\\
\end{center}
\baselineskip=12pt
{\bf{Table 1.}} {\small
Dependence of $\alpha_3(M_Z)$ on $\sin^2 \theta_W(M_Z)$
and $T_{SUSY}$, in the framework of  gauge
coupling unification.}\\
\baselineskip=18pt

The above values of $\alpha_3(M_Z)$
are obtained from our two--loop renormalization
group analysis, and  they depend
slightly  on the top
quark Yukawa coupling, and hence on $\tan\beta$. In general, for
values of the top quark mass $M_t \geq 140$ GeV, the variation induced
through the top quark Yukawa coupling contribution
 is at most 1$\%$ of the values given above
(with lower values of $\alpha_3(M_Z)$ obtained for larger values
of the top quark Yukawa coupling).
Observe that in models with universal gaugino masses at the
grand unification scale and for large values of the supersymmetric
mass parameters, for which the mixing in the neutralino and
chargino sectors may be neglected, $T_{SUSY} \simeq \mu/6$,
where $\mu$ is the supersymmetric mass parameter appearing in the
superpotential. Hence, a value of $T_{SUSY}$ of the order of
1 TeV implies that the supersymmetric spectrum contains sparticles
with masses far above the TeV scale. If all supersymmetric masses
are below or of the order of 1 TeV, the effective supersymmetric
threshold scale is below or of the order of the weak scale.

The largest uncertainties associated with the unification scheme
come from the threshold corrections at the grand unification
scale.  We shall not discuss them here (for a detailed discussion,
see, for example, Ref. \cite{LP}), but we shall assume moderate
corrections, of the order of
those  coming from the supersymmetric spectrum. Thus,
for a supersymmetric spectrum with characteristic masses of the
order of or below 1 TeV, the condition of gauge coupling
unification, together with the experimental correlation between
$\sin^2\theta_W(M_Z)$ and $M_t$,
Eq. (\ref{eq:sint}), imply the following
correlation between $\alpha_3(M_Z)$ and the top quark mass,
\begin{equation}
M_t^2 \simeq (138 \;
 {\mathrm{ GeV }})^2 + 0.25 \times \; 10^7 \; {\mathrm{ GeV}}^2 \;
\left( \alpha_3(M_Z) - 0.123 \pm 0.01 \right).
\label{eq:al3tm}
\end{equation}
It is instructive to compare Eq. (\ref{eq:al3tm}) with the results
presented in table 1. As we discussed in section 1, a lower bound
for $\alpha_3(M_Z)$ as a function of the top quark mass
is derived. From Eq. (\ref{eq:al3tm}) it follows that, for
a top quark mass $M_t \geq 150 \; (170)$ GeV, the values of the
strong gauge coupling are
$\alpha_3(M_Z) \geq 0.114 \; (0.117)$.

\section{Yukawa Coupling Unification}

The general features of Yukawa coupling unification in the Minimal
Supersymmetric Standard Model are discussed in section 1.
In the following, we shall concentrate on the properties of the
minimal supersymmetric SO(10) model, for which not only the
bottom and the tau, but also the top quark Yukawa coupling
unify at the GUT scale. As we discussed in section 1, large
values of $\tan\beta$ $(\simeq {\cal{O}}(50)$), are predicted
within this scheme.

The fact that in the minimal supersymmetric SO(10) model,
 the value of $\tan\beta$ is approximately equal to
the ratio of the top and bottom quark masses at the top mass
scale comes from the approximate equality of the top and bottom
Yukawa couplings at low energies. Such an approximate equality is
implied by their unification condition, and the
presumption that the bottom and top quarks acquire masses, each of them,
only through one of the two Higgs doublet vacuum expectation values:
$m_t(m_t) = h_t(m_t) v_2$, $m_b(m_t) = h_b(m_t) v_1$, where
$v_i$ is the vacuum expectation value of the Higgs $H_i$.
This property holds in the supersymmetric limit and,
within a good approximation, for small and moderate values
of $\tan\beta$ when supersymmetry is softly broken. In general, however,
a coupling of the bottom (top) quark to the Higgs
$H_2$ ($H_1$) may be generated at the one--loop level. Although
these couplings are small compared to $h_b$ $(h_t$)
 (typically lower than
1$\%$ of $h_b$), for large values of $\tan \beta$ --which implies
 $v_2 \gg v_1$-- this may induce important
 corrections to the bottom mass \cite{Banks}--\cite{Ralf}, \cite{wefour}:
\begin{equation}
m_b = h_b (v_1 + K_1 v_2) \equiv \tilde{m}_b
\left( 1 + \frac{ \Delta m_b }{ \tilde{m}_b } \right)  ,
\label{eq:Deltamb}
\end{equation}
where $K_1$ is the coefficient of the one--loop corrections to
the bottom mass, $\Delta m_b/\tilde{m}_b = K_1 \tan\beta$, and
$\tilde{m}_b$ would be the value of the running bottom mass if
the supersymmetric one--loop corrections were negligible.
Recalling that, at the two--loop level, the physical and
running bottom masses are related by
\begin{equation}
M_b = m_b(M_b) \left( 1 + \frac{4}{3\pi} \alpha_3(M_b) +
12.4 \left(\frac{\alpha_3(M_b)}{\pi} \right)^2 \right),
\end{equation}
due to the low energy renormalization group running of the
bottom quark mass
the following property is fulfilled,
\begin{equation}
\frac{\Delta M_b}{ \tilde{M}_b } \simeq
\frac{ \Delta m_b(M_b) }{ \tilde{m}_b(M_b) } \simeq
\frac{ \Delta m_b(M_Z) }{ \tilde{m}_b(M_Z) }.
\label{eq:relbm}
\end{equation}
In the above, $\tilde{M}_b$ would be the value of the
physical bottom mass if no supersymmetric corrections were
present, while $\Delta M_b \simeq M_b - \tilde{M}_b$.

The above property, Eq. (\ref{eq:relbm}), is relevant for the
understanding of the top quark mass predictions coming from
the unification of couplings. In Fig. 1,
we present the predictions
for the top quark mass as a function of $\alpha_3(M_Z)$ for
different values of the {\it uncorrected} bottom quark
mass $\tilde{M}_b$, as well as for different values of
$\tan\beta$ \cite{wefour}.
%
%
%
%
%
%
%
%
%
%
%
{}From this figure  we observe that the top quark mass predictions are
strongly dependent on the exact value of the strong gauge coupling
and the {\it uncorrected} bottom mass $\tilde{M}_b$. Notice that,
if we restrict ourselves to the region of $\alpha_3(M_Z)$
preferred by gauge coupling unification, Eq. (\ref{eq:al3tm})
(to the right of the dotted line), for $M_{\tilde{b}} \leq$
5 GeV, the top quark mass is pushed towards large values.
 In fact, if the
bottom mass corrections were small, $\tilde{M}_b \simeq M_b$,
for experimentally acceptable values of the bottom quark pole
mass, $M_b = 4.9 \pm 0.3$ GeV \cite{Pdat},
the top quark mass would be
 $M_t \geq 165$ GeV.  However, as we shall show in section 5,
large bottom quark mass corrections may significantly change
this prediction.

\vspace*{8cm}
{}~\\
\baselineskip=10pt
{\small Fig. 1. Top quark mass predictions as a function of the
strong gauge coupling $\alpha_3(M_Z)$, with unification of the three
Yukawa couplings of the third generation.  Starting from above,
the solid lines represent
values of  $\tilde{M}_b$  equal to 4.6, 4.9, 5.2, 5.5 and 5.8 GeV.
Analogously, the dashed lines represent values of
 $\tan\beta$ equal to 60, 55, 50, 45 and 40. The long-dashed line
represents the top quark mass fixed point value and the dotted line
shows the region preferred by the gauge coupling unification
condition as explained in the text.}
{}~\\
\baselineskip=18pt

For  a clear interpretation of Fig. 1 it is important that, for
large values of $\tan\beta$, the fixed point prediction,
Eq. (\ref{eq:fixmt}), is modified due to the non-negligible
bottom quark Yukawa coupling effect in the running of the top quark
Yukawa coupling \cite{CPW},\cite{wefour}.
Indeed, for $h_b \simeq h_t$ a more appropriate expression
than Eq. (\ref{eq:fixmt}) is

\begin{equation}
m_t(M_t)^{IR} \simeq 182 \; {\mathrm{ GeV}}
\left[1 + 2 \; (\alpha_3(M_Z) - 0.12)\right]
\label{eq:fplar}
\end{equation}
which, recalling the relation between the running and the
pole masses, Eq. (\ref{eq:runpomt}),
describes within a good approximation the
upper bound on the top quark mass (long-dashed line)
shown in Fig. 1.

\section{Supersymmetry and Electroweak Symmetry Breaking}

As we mentioned above, in the large $\tan\beta$ regime, the
top quark mass predictions depend strongly on the nature of
the soft supersymmetry breaking terms arising at low energies.
In principle, the Minimal Supersymmetric Standard Model
yields   a multiplication of free parameters, which are
only constrained by the requirement of avoiding a  conflict
with present experimental data. One of the strongest requirements
is the absence of flavour changing neutral
currents, which is  fulfilled if the squarks of the first
two generations are approximately degenerate in mass.
It has been realized long ago that such is
naturally the case if all soft supersymmetry breaking
squark and gaugino mass parameters
are universal at the grand unification scale.
This
supersymmetry breaking scheme appears naturally in minimal
supergravity models, in which not only the squarks
but also the Higgs fields acquire common soft supersymmetry
breaking terms at high energies, and all the supersymmetry
breaking terms at low energies may be given as a function of
only four parameters: The universal scalar mass $m_0$, the
universal gaugino mass $M_{1/2}$, and the universal trilinear
and bilinear couplings $A_0$ and $B_0$, which are  associated
with contributions to the high energy
effective potential proportional to the trilinear and
bilinear terms of the superpotential.

We shall first concentrate on the minimal supergravity SO(10)
model. In  section 6 we shall briefly discuss the situation where
the condition of universality of the soft supersymmetry
breaking parameters
is relaxed. In general, the superpotential reads
\begin{eqnarray}
P  =  \mu \epsilon_{ab} H_1^a H_2^b + h_t \epsilon_{ab}
Q^a U H_2^b
 +  h_{\tau} \epsilon_{ab} L^a E H_1^b + h_b \epsilon_{ab}
Q^a D H_1^b  \; ,
\end{eqnarray}
where $Q^T = (T_L, B_L)$ and  $L^T = (\nu^{\tau}_L, \tau_L)$ are
the top--bottom and lepton left--handed doublets,
$U = T_L^c$, $D = B_L^c$ and $E = \tau^c_L$ are the right--handed
top--squark, bottom--squark and stau, respectively,
 and $\epsilon_{ab}$
is the antisymmetric tensor with $\epsilon_{12} = -1$. The scalar
potential is given by
\begin{eqnarray}
V &=& m_1^2 H_1^{\dagger} H_1 + m_2^2 H_2^{\dagger} H_2 +
m_Q^2 Q^{\dagger} Q + m_U^2 U^* U  + m_D^2 D^* D +
m_L^2 L^{\dagger} L + m_E^2 E^* E
\nonumber\\
 &+& \epsilon_{ab} (m_3^2  H_1^a H_2^b +
A_t h_t Q^a U H_2^b + A_b h_b Q^a D H_1^b + {\mathrm{ h.c.}}) +
{\mathrm{ q.t.}} + ... \; ,
\end{eqnarray}
where the Higgs field mass terms contain a part coming
from the superpotential and another coming from the
soft supersymmetry breaking parameters: $m_i^2 = \mu^2 +
m_{H_i}^2$, with $i=1,2$ and   $m_3^2 = B \mu$.
 The dots stand for scalar trilinear {\it F}-terms
 and q.t.
characterizes the quartic terms in the scalar fields. The
soft supersymmetry breaking parameters $m_{H_i}^2$ have
the following renormalization group evolution,

\begin{eqnarray}
4 \pi \frac{ d m_{H_1}^2 }{ dt} &=&
3 \alpha_2 M_2^2 + \alpha_1 M_1^2 - 3 Y_b M_{Deff}^2
- Y_{\tau} M_{Eeff}^2,
\nonumber\\
4 \pi \frac{ d m_{H_2}^2 }{ dt} &=&
3 \alpha_2 M_2^2 + \alpha_1 M_1^2 - 3 Y_t M_{Ueff}^2,
\label{eq:m12}
\end{eqnarray}
where $t = \ln[(M_{GUT}/Q)^2]$, $Y_i = h_i^2/(4\pi)$,
$M_{Deff}^2 = m_Q^2 + m_D^2 + m_{H_1}^2 + A_b^2$,
$M_{Ueff}^2 = m_Q^2 + m_U^2 + m_{H_2}^2 + A_t^2$ and
$M_{Eeff}^2 = m_L^2 + m_E^2 + m_{H_1}^2 + A_{\tau}^2$.

The rest of the renormalization group equations may be found
in the literature \cite{Savoy}.
Let us just remark that, in the minimal
supergravity scheme, $M_{D eff}$ and $M_{U eff}$  present
very similar renormalization group evolutions  when
$h_t \simeq h_b$.  Indeed,
for bottom--top Yukawa coupling unification
they only differ by the different hypercharge quantum numbers
of the right bottom and top quarks, the
slightly different running of the
bottom and top Yukawa couplings, and the
small tau Yukawa coupling effects (recall that the tau Yukawa
coupling is  renormalized to lower values than the
bottom and top ones, due to the absence of strong gauge coupling
effects in its  one--loop renormalization group evolution).

Considering the renormalization group evolution of the mass
parameters $m_{H_1}^2$ and $m_{H_2}^2$, for $M_{1/2}^2
\gg m_0^2$, values of $m_{H_1}^2 > m_{H_2}^2$ are obtained,
mainly due to the difference between
the hypercharge quantum numbers
of the right bottom and top quarks and their supersymmetric
partners.  For $m_0^2 \gg M_{1/2}^2$, instead,
the inverse hierarchy, $m_{H_2}^2 > m_{H_1}^2$, is obtained,
mainly due to the $\tau$-Yukawa coupling effects.
In general,
considering the bottom--top Yukawa coupling unification condition and
performing a complete numerical analysis
it follows that \cite{wefour}
\begin{equation}
m_{H_1}^2 - m_{H_2}^2 \simeq \alpha M_{1/2}^2 + \beta m_0^2,
\label{eq:diff12}
\end{equation}
with $\alpha \simeq - \beta \simeq 0.1$ -- $0.2$,
depending on the proximity
of the top quark Yukawa coupling to its infrared fixed point value
$h_f$ (the above range is obtained for
$Y_t/Y_f = h_t^2/h_f^2 = $ 0.6 -- 0.95, respectively).

\subsection{Radiative Electroweak Symmetry Breaking}

In order to induce a proper breakdown of the electroweak symmetry, the
following conditions need to be fulfilled,
\begin{equation}
\sin 2 \beta = \frac{ 2 m_3^2}{m_A^2}
\end{equation}
and
\begin{equation}
\tan^2 \beta =
\frac{ m_1^2 + M_Z^2/2 + {\mathrm{ r.c.}}}{m_2^2 +
 M_Z^2/2 + {\mathrm{ r.c.}}} ,
\end{equation}
where $m_A^2 \simeq m_1^2 + m_2^2 +$ r.c. is the mass of the CP-odd
Higgs eigenstate and
r.c. simbolizes one--loop radiative correction contributions,
which depend logarithmically on  the supersymmetry breaking scale,
and become of the order of $M_Z^2$ for a characteristic
supersymmetric scale of the order of 1 TeV. We shall ignore these
corrections in the following, since they are unessential
for the qualitative understanding of the phenomena under discussion.
They are included, however, in the numerical analysis.

In general, independently of the supersymmetry breaking mechanism,
since $\tan\beta$ is very large, in order to avoid extremely large
values of $m_1^2$ (or $m_A^2$), the mass parameters $m_2^2$ and
$m_3^2$ should fulfil the following properties
\begin{equation}
m_2^2 \simeq - \frac{M_Z^2}{2} , \;\;\;\;\;\;\;\;\;\;\;\;\;\;
m_3^2 \ll M_Z^2.
\label{eq:ewcond}
\end{equation}

As was  explained in Ref. \cite{Hall}, the second of these
conditions can be obtained by assuming that there is a softly
broken symmetry implying the smallness of the parameters
$B$ and/or $\mu$. The first condition is just a reflection of
the fact that the vacuum expectation value of the Higgs $H_2$
is the one that determines the electroweak vector boson masses.
The fact that the CP-odd Higgs mass squared should be positive,
together with Eq. (\ref{eq:ewcond}), implies that
\begin{equation}
m_1^2 - m_2^2 > M_Z^2 \; .
\label{eq:diff}
\end{equation}

The above described properties are general in the sense that
they do not depend on the supersymmetry breaking mechanism.
If we consider the  running of the Higgs mass parameters in
the minimal supergravity model with bottom--top Yukawa unification,
Eq. (\ref{eq:diff12}), strong implications follow from
Eqs. (\ref{eq:ewcond}) and  (\ref{eq:diff}),
\begin{equation}
M_{1/2} > m_0, \;\;\;\;\;\;\;\;\;\;\;\; M_{1/2} \geq
\frac{M_Z}{\sqrt{\alpha}} .
\label{eq:21}
\end{equation}
Analysing the RG evolution of the soft
supersymmetry breaking parameters,
Eq. (\ref{eq:m12}), and taking into account the constraints of
Eq. (\ref{eq:21}), an approximate solution
for the Higgs mass parameter $m_2^2$ is  obtained,
\begin{equation}
m_2^2 \simeq \mu^2 + M_{1/2}^2 \left( 0.5 - C_1 \frac{Y}{Y_f}
+ C_2 \left(\frac{Y}{Y_f}\right)^2 \right) ,
\label{eq:mum12}
\end{equation}
where $Y/Y_f$ is the ratio of the top quark Yukawa coupling
squared to its fixed point value, and $C_1$ and $C_2$ are
coefficients that depend on the value of the strong gauge
coupling constant and,
for $\alpha_3(M_Z) \simeq 0.12$,
are approximately given by $C_1 \simeq 6$
and $C_2 \simeq 3$.
Hence, the condition $m_2^2 \simeq -M_Z^2/2$, together with
the large values of $M_{1/2}$ necessary to achieve unification
of couplings, imply a strong correlation between $\mu$ and
$M_{1/2}$. This  correlation, which is  shown in Fig. 2,
has  profound implications for the sparticle spectrum and
the determination of the bottom mass corrections.

In the following, we
shall summarize the main features of the Higgs and supersymmetric
spectrum. For a more detailed discussion, we  refer the reader
to  \cite{wefour}.
%
%
%
%
\vspace*{10cm}
{}~\\
 {\small Fig. 2.
Supersymmetric mass parameter $\mu$ as a function of
the gaugino mass $M_{1/2}$ for two different values of the
top quark mass in the framework of bottom--top Yukawa unification.
Only the lower top quark mass
value $M_t \simeq 150$ GeV leads to an acceptable value of
the physical bottom mass, once the supersymmetry breaking-induced
bottom mass corrections are included.}\\

The  main features of the
sparticle spectrum are governed by the
three properties explained above: a) Strong $\mu$--$M_{1/2}$
correlation, b) Large values of $M_{1/2} \geq 300$ GeV,  and
c) $M_{1/2} \geq m_0$ . This yields:\\
1) Small mixing in the chargino and neutralino sectors, which
naturally follows from properties a) and b).
The lightest supersymmetric
particle is mainly a bino, with mass $M_{\tilde{B}} \simeq 0.4 M_{1/2}$.
The second--lightest neutralino and the lightest chargino are winos
and hence almost degenerate in mass $M_{\tilde{W}^+} \simeq
M_{\tilde{W}^0} \simeq 2 M_{\tilde{B}}$. The heaviest neutralino
and chargino are  Dirac (pseudo--Dirac in the neutralino case)
particles, with masses approximately equal to the parameter $|\mu|$.
{}~\\
2) Strong stop (sbottom)--gluino running
mass correlations,
with $M_{\tilde{t}} \simeq
M_{\tilde{b}} \simeq 0.75$--$0.8 M_{\tilde{g}}$
 $ (0.85$--$0.9 M_{\tilde{g}}$)
for the
lightest (heaviest) squark mass eigenstate. The
running gluino mass is related to the common gaugino mass by
$M_{\tilde{g}} \simeq 2.6$--$2.8  \; M_{1/2}$.\\
3) Large mixing in the stau sector, due to the off--diagonal
left--right stau matrix element $m^2_{\tilde{\tau}_{LR}}
 \simeq - h_{\tau} \mu v_2
\simeq - m_t \mu/\sqrt{3}$.  The large stau mixing leads to a
lower bound on $m_0$,
\begin{equation}
m_0^2 \geq -0.15 M_{1/2}^2 + \left( (0.15 M_{1/2}^2 )^2  +
\mu^2 m_t^2/3 \right)^{1/2}
\end{equation}
in order to avoid a stau being the lightest supersymmetric particle.\\
4) The Higgs spectrum is characterized by
relatively low values for the
CP-odd Higgs mass,  $ m_A \leq M_{\tilde{t}} \; (\alpha/5)^{1/2}$,
where $\alpha$ is the coefficient characterizing the dependence
of $m_{H_1}^2 - m_{H_2}^2$ on $M_{1/2}^2$, Eq. (\ref{eq:diff12}).
Observe that, due to the characteristic values of
$\alpha \simeq 0.2 \; (0.1)$ for $M_t \simeq 190 \; (160)$ GeV,
if the squark masses are lower than a few TeV, the CP-odd and
charged Higgs masses will be lower than a few hundred GeV.\\
5) The lightest CP-even scalar would
become lighter than the Z-boson
if the CP-odd Higgs scalar were light, $m_A \leq
150$ GeV.  However,
in the minimal supergravity SO(10) model, such low values of
$m_A$ induce large values of the $b \rightarrow s \gamma$ decay
rate, since
the chargino contribution  enhances the rate with respect
to that of the Standard Model with an extra Higgs doublet,
if the correct bottom mass predictions are required.
Therefore, in the minimal supergravity SO(10)
model the lightest CP-even Higgs mass is larger than the Z
boson mass. We shall come back to this issue in  section 5.

\section{Bottom Mass Predictions and the $b \rightarrow s \gamma$
Decay Rate}

In section 3 we have mentioned the possibility of inducing
non--negligible
one--loop corrections to the running bottom mass,
Eq. (\ref{eq:Deltamb}). In this section we want to  explicitly show which
are the relevant one--loop contributions to the running bottom
mass and how  they are correlated with the chargino contributions to
the $b \rightarrow s \gamma$ decay rate.
As we show in section 3,
in order to estimate the size of the bottom mass corrections, we
need to compute the supersymmetry breaking-induced effective
coupling of the bottom quark to the $H_2$ field, which amounts to
computing the coefficient $K_1$, with $\Delta m_b/\tilde{m}_b =
K_1 \tan\beta$. The contributions to the coefficient $K_1$ come
from the  sbottom--gluino  and stop--chargino graph contributions
to the bottom quark self energies. It is easy to show that the
dominant contributions are given by
\begin{equation}
K_1 = \frac{2 \alpha_3}{3 \pi} \; M_{\tilde{g}} \mu \;
I(m^2_{\tilde{b}_1},m^2_{\tilde{b}_2},M_{\tilde{g}}^2) +
\frac{Y_t}{4\pi} \; A_t \mu \;
I(m^2_{\tilde{t}_{1}},m^2_{\tilde{t}_{2}},\mu^2)  \; ,
\end{equation}
where $m^2_{\tilde{q}_{i}}$,  with $i=1,2$, are the squark mass
eigenstates and the integral factor $I(a,b,c)$ is given by
\begin{equation}
I(a,b,c) = \frac{ a b \log(a/b) + b c \log(b/c) + a c \log(c/a)}
{(a - b) (b - c) (a - c)} .
\end{equation}
The renormalization group equation  of the $A_t$ parameter shows
that its low energy values are strongly correlated with the
universal gaugino mass. In fact, using the relation $M_{\tilde{g}}
\simeq$ 2.6--2.8 $M_{1/2}$ it follows \cite{wefour} that

\begin{eqnarray}
A_t \simeq  - \frac{2}{3} M_{\tilde{g}}, \;\;\;\;\;\;\;\;\;
{\mathrm{ for}} \;\;\; Y/Y_t \simeq 1
\nonumber\\
A_t \simeq - M_{\tilde{g}} \;\;\;\;\;\;\;\;\;\; {\mathrm{ for}} \;\;\;
Y/Y_f \simeq 0.6.
\label{eq:At}
\end{eqnarray}
Observe that, due to the minus sign appearing in Eq. (\ref{eq:At}),
there is a partial cancellation between the two different contributions
to $K_1$, which yields a significant reduction in the bottom mass
corrections (typically of the order of 25$\%$). As will be shown
below, this partial cancellation, although important, is by far not
sufficient to render the bottom mass corrections small.

\subsection{Conditions for a small $K_1$ and Minimal Supergravity}

To get an estimate of the size of the bottom mass corrections,
it is important to observe that the integral factors $I(a,b,c)$
are always of the order of the  inverse of the
largest mass squared appearing
in the integral.  Since the coupling constant dependent
factors of both contributions to $K_1$ are of order 0.01, and
$\tan\beta$ is of order 50,  to get a small
bottom mass correction the following
properties should be fulfilled \cite{Hall}:
\begin{equation}
M_{\tilde{g}} \ll m_{\tilde{q}} ,  \;\;\;\;\;\;\;\;\;\;\;
{\mathrm{ and/or}}     \;\;\;\;\;\;\;\;\;\;\;\;\;
\mu \ll m_{\tilde{q}},
\end{equation}
where $m_{\tilde{q}}$ represents the heaviest
third generation squark mass
eigenstates. In addition, the requirement of electroweak symmetry
breaking implies that the mass parameter $B$ should
be small in comparison to $m_A$, unless $\mu$ itself is much
smaller than $m_A$. All these requirements may be satisfied by
imposing a softly broken Peccei-Quinn symmetry, which implies
the smallness of the mass parameter $\mu$, together with an
approximate continuous $R$ symmetry, present in the limit
$B \rightarrow 0$, $M_{\tilde{g}} \rightarrow 0$, $A_{t}
\rightarrow 0$, whose breaking is characterized by the (assumed)
small parameter
\begin{equation}
\epsilon_R \simeq
\frac{B}{m_{\tilde{q}}} \simeq
\frac{A_t}{m_{\tilde{q}}} \simeq
\frac{M_{\tilde{g}}}{m_{\tilde{q}}}   \; .
\label{eq:Rsymm}
\end{equation}
As we shall discuss in  section 6,  the above  conditions may only be
reached by relaxing the  condition of universality of  the
soft supersymmetry breaking parameters at the grand unification
scale.

In the framework of minimal supergravity,
with exact unification of the third generation quark and lepton
Yukawa couplings,
the strong correlations between
the parameters $\mu$, $A_t$, $M_{\tilde{g}}$
and the third
generation squark masses
--derived from their renormalization group equations
and the condition  of a proper
breakdown of the electroweak symmetry-- imply that the above
symmetries are not
present in the low energy spectrum. Hence,
the bottom mass
corrections are  large within this framework \cite{wefour}:
\begin{equation}
\frac{\Delta m_b}{\tilde{m_b}}
 \simeq 0.0045 \tan\beta \left( \frac{\mu}{M_{1/2}}
\right).
\end{equation}
Taking into account Eq. (\ref{eq:mum12}) and
the numerical results from Fig. 1, we see that the
corrections are of order 45 $\%$ for $M_t \simeq 190$ GeV
and of order 20 $\%$ for $M_t \simeq 150$ GeV.

The corrections are sufficiently large to rule out any
solution with $\tilde{M}_b < M_b$.  This is simply due to
the impossibility
to accommodate a physical bottom mass in the experimentally
allowed range, while keeping the top Yukawa coupling in
the perturbative domain at energies of the order of the
grand unification scale. Hence, the coefficient $K_1$
should be negative, implying that the only acceptable branch
is that  with negative (positive) values of $M_{\tilde{g}}
\times \mu$ ($A_t \times \mu$). Consequently, an upper bound on
the top quark mass $M_t$ may be obtained.
This corresponds, for a given value of the strong gauge coupling,
to the maximum
value of  the top quark Yukawa coupling
(and $\tan \beta$)  consistent with a
 physical bottom mass $M_b$  equal to its
lower experimental bound $M_b^L \simeq 4.6$ GeV. There are
small uncertainties in the computation of this upper bound,
associated with the size of the low energy threshold corrections
to the top quark mass, small tau mass corrections analogous
to the bottom ones  and QCD scale uncertainties. Conservative
upper bounds for the top quark mass are given by \cite{wefour}
\begin{eqnarray}
M_t \leq 165 \; (175) \; (185) \; {\mathrm{ GeV}}, \;\;\;\;\;\;\;\;\;
{\mathrm{ for}} \;\;\;\; \alpha_3(M_Z) = 0.12 \; (0.125) \; (0.13).
\end{eqnarray}
These bounds go rapidly down if the bottom mass is larger
than 4.6 GeV. Let us remark again that these bounds do not
apply in  general in the supersymmetric
 SO(10) model, but are only a
consequence of the particular supersymmetry breaking
scheme under study. Relaxing the high energy boundary  conditions for
the soft supersymmetry breaking parameters, these bounds on $M_t$ may be
diluted.

\subsection{ $b \rightarrow s \gamma$ Decay Rate}

The dominant supersymmetric contributions to the $b \rightarrow
s \gamma$ decay rate have been recently analysed by several
authors \cite{BBMR}--\cite{Stefano}.
In the Minimal Supersymmetric Standard Model, there is a
contribution coming from  the charged Higgs, which, for low
values of the charged Higgs mass,  enhances the Standard
Model decay rate. The dominant effect from
supersymmetric particles
comes from the one--loop chargino--stop contributions to the
$b_R \rightarrow s_L \gamma$ transition. In the supersymmetric limit,
$\tan\beta = 1$ and $\mu = 0$, the chargino contributions exactly
cancel the $W^{\pm}$ and charged Higgs ones, and the $b \rightarrow
s \gamma$ transition element vanishes.
This behaviour is not  preserved once supersymmetry is broken.

In particular, the large $\tan\beta$ scenario is far from being
close to the supersymmetric limit and the dominant chargino
contribution to the $b \rightarrow s \gamma$ decay rate may
have both signs. In general,
in the large $\tan\beta$ regime it is proportional to
\begin{equation}
A_{\gamma,g} \simeq \frac{m_t^2}{m_{\tilde{t}}^2} \;
\frac{A_t \mu}{m_{\tilde{t}}^2}   \; \tan \beta
\; g_{\gamma,g} \left(\frac{m_{\tilde{t}}^2}{\mu^2} \right) ,
\label{eq:AGG}
\end{equation}
where $A_{\gamma}$ and $A_{g}$ are the coefficients of the
effective operators for $b s$--photon  and $b s$--gluon
interactions, as defined in Ref. \cite{BG},
$g_{\gamma,g}(x)$ is  a
function of $x$ proportional to the derivative of  the function
$f^{(3)}_{\gamma,g}(x)$ defined in Ref. \cite{BG}, and we
have assumed a small mixing in the stop sector, with
mass eigenstates $m_{\tilde{t}_{1(2)}} = m_{\tilde{t}}^2
+(-) A_t m_t$.
{}From Eq. (\ref{eq:AGG}) it follows that
the sign of the chargino contribution
to the $b \rightarrow s \gamma$ decay amplitude depends
on the sign of $A_t \times \mu$, and hence is correlated in sign
with the bottom mass corrections discussed above. Observe that
the chargino (charged Higgs) contribution to the decay
amplitude is always small if the supersymmetric (charged
Higgs) spectrum is sufficiently heavy.

One can show that for
positive (negative) values of  $A_t \times \mu$ the
supersymmetric rate becomes larger (smaller) than that of the
Standard Model plus one extra Higgs doublet. Since in the minimal
supergravity SO(10)
model, once the supersymmetry breaking-induced bottom mass corrections
are included, positive values of $A_t \times \mu$ are
required to obtain acceptable
values for the physical bottom mass, then,  the
 $b \rightarrow s \gamma$ decay rate may be significantly enhanced
with respect to the SM one.
Consequently, one can obtain sparticle
mass bounds by requiring the decay rate to be smaller than the
present experimental bounds, $BR(b \rightarrow s \gamma) <
5.4 \; \times 10^{-4}$ \cite{BSGAB}.
One should recall, however, that there are
theoretical uncertainties associated with the decay rate
computations, which may be as large as $20$--$30\%$ \cite{PB}. Taking
them into account at the 2--$\sigma$ level \cite{wefour},
for a top quark mass $M_t$ of the order of
150 GeV,
the following lower bound on the universal mass parameter $M_{1/2}$ may
be obtained:
\begin{equation}
M_{1/2} \geq 550\; {\mathrm{ GeV}}.
\end{equation}
This implies a lower bound for the masses
of the third generation
squarks, charginos and charged Higgses larger than the
ones already demanded by the condition of electroweak
symmetry breaking,
\begin{equation}
m_{\tilde{q}} \geq 1.2 \; {\mathrm{ TeV}}, \;\;\;\;
m_{\tilde{\chi^+}} \geq 450 \; {\mathrm{ GeV}}, \;\;\;\; m_{H^+} \geq
180 \; {\mathrm{ GeV}}.
\end{equation}
Observe that, in this case, the lightest
CP-even Higgs mass is constrained to be in the
range $m_h \simeq 110$--$130$ GeV.

\section{Non-Universal Soft Supersymmetry Breaking Parameters}

In the above, we have analysed in detail the implications
of having universal soft supersymmetry breaking parameters at
the grand unification scale. In this case, the bottom mass
corrections and the chargino contributions to the
$b \rightarrow s \gamma$ decay rate are correlated in sign and
they are sufficiently
large to put constraints on the particle spectrum as well as
an upper bound on the top quark mass as a function of the strong
gauge coupling. The phenomenological implications
of relaxing the condition of universality
depend on the nature of the soft supersymmetry
breaking parameters. Instead
of performing a detailed study of the general case,
we shall concentrate
on understanding which should be
the pattern of  supersymmetry breaking parameters at the
grand unification scale in order to induce only small bottom
mass corrections (lower than, say, 10$\%$ for a heavy top quark)
and small chargino contributions to the $b \rightarrow s \gamma$
decay rate.

We assume that the universality of the
soft supersymmetry breaking gaugino masses is kept and we
investigate  possible violations of the universality
condition in the scalar sector. As  was shown in section 5,
to obtain an effective cancellation of the bottom quark mass
corrections and of the chargino contributions to the
$b \rightarrow s \gamma$ decay rate,
we need either small values of $A_t$ and
$M_{\tilde{g}}$ and/or
small values of $\mu$ in comparison with the squark masses.
Hence, the soft supersymmetry
breaking parameters should fulfil  the following property
\begin{equation}
A_{t,b}(0) \simeq B(0) \simeq M_{1/2} \ll m_S(0),
\label{eq:domsc}
\end{equation}
where $m_S(0)$ represents the characteristic scalar soft
supersymmetry breaking mass parameters.
Assuming bottom--top--tau Yukawa coupling unification and
the fulfilment of Eq. (\ref{eq:domsc}),
and
neglecting the small tau Yukawa coupling
effects in the bottom quark Yukawa coupling renormalization
group running as well as the small right top--bottom hypercharge
difference, the following approximate analytical solutions
for the Yukawa couplings and mass parameters are obtained,
\begin{eqnarray}
m_{H_2}^2 & = & m_{H_2}^2(0) - 3 I_U, \;\;\;\;\;\;\;\;\;\;\;\;\;\;\;
m_{H_1}^2  =  m_{H_1}^2(0) - 3 I_D - I_L,
\nonumber\\
m_{Q}^2 & = & m_{Q}^2(0) - I_U - I_D, \;\;\;\;\;\;\;\;\;\;
m_{U}^2   =  m_{U}^2(0) - 2 I_U,
\nonumber\\
m_D^2 & = & m_D^2(0) - 2 I_D,    \;\;\;\;\;\;\;\;\;\;\;\;\;\;\;\;
Y_b \simeq Y_t  =  \frac{4 \pi Y(0) E}{ 4 \pi + 7 Y(0) F},
\end{eqnarray}
where
\begin{eqnarray}
I_U & = & \frac{Y}{7\;Y_f} \left( m_{H_2}^2(0) + m_Q^2(0)
+ m_U^2(0) \right)
\nonumber\\
& + & \frac{1}{10} \left(
m_{H_1}^2(0) - m_{H_2}^2(0) + m_D^2(0) - m_U^2(0) \right)
\left[  \frac{5 Y}{7 Y_f} + \left( 1 - \frac{Y}{Y_f}
\right)^{5/7}  - 1 \right] ,
\nonumber\\
I_D & = & \frac{Y}{7\;Y_f} \left( m_{H_1}^2(0) + m_Q^2(0)
+ m_D^2(0) \right)
\nonumber\\
& - & \frac{1}{10} \left(
m_{H_1}^2(0) - m_{H_2}^2(0) + m_D^2(0) - m_U^2(0) \right)
\left[  \frac{5 Y}{7 Y_f} + \left( 1 - \frac{Y}{Y_f}
\right)^{5/7} - 1 \right] ,
\nonumber\\
I_L & \simeq & \frac{1}{6} (m_{H_1}^2(0) + m_L^2(0) + m_E^2(0) )
\left[ 1 - \left( 1 - \frac{Y}{Y_f} \right)^{1/4} \right].
\end{eqnarray}
In the above,
$Y = Y_t \simeq Y_b$, with $Y_i = h_i^2/4\pi$,
$E$ and $F$ are functions of the
gauge couplings --which at the weak scale take values
$E \simeq 14$ and $F \simeq 290$--
and $Y_f$ is the fixed point value for the top quark Yukawa coupling,
$Y_f \simeq 4 \pi E/7 F$,
obtained whenever $Y(0) \geq 0.1$ and
whose associated mass expression at the
two--loop level
is given in Eq. (\ref{eq:fplar}). The functions $I_U$ and
$I_D$ are obtained from an exact solution to the coupled renormalization
group equations in the limit of negligible tau
Yukawa coupling effects. The function $I_L$ provides
a good approximation to the Yukawa coupling effect in the range
$Y/Y_f \simeq 0.6$--0.95, and has been obtained assuming that $Y_{\tau}$
rapidly acquires its low energy hierarchy with
respect to $Y_b$ and $Y_t$.

The interrelation among the different Higgs, squark and slepton
masses is quite relevant. For instance, even if we start with
values of $m_{H_2}^2(0)$ lower than $m_{H_1}^2(0)$ at the
grand unification scale, depending on the relation between the
squark mass parameters $m_{U}^2(0)$ and $m_{D}^2(0)$, one may
or may not find an acceptable solution.
Close to the infrared fixed point
of the  top quark mass, Eq. (\ref{eq:fplar}), the following
relations
\begin{eqnarray}
m_Q^2 + m_U^2 + m_{H_2}^2 \simeq 0
\nonumber\\
m_Q^2 + m_D^2 + m_{H_1}^2 \simeq 0
\end{eqnarray}
are fulfilled,
implying a strong correlation between the low energy
values of the Higgs and squark mass parameters. In particular,
for large values of the squark masses $m_{\tilde{q}}^2 \gg
M_Z^2$, since $m_2^2 \simeq -M_Z^2/2$,  the above yields:
\begin{equation}
\mu^2 \simeq -m_{H_2}^2 \simeq m_Q^2 + m_U^2.
\end{equation}
Hence, for values of the top quark mass close
to its fixed point value, the bottom mass corrections and
the chargino contributions to the $b \rightarrow s \gamma$
decay rate will only be suppressed if $M_{\tilde{g}}$ is
much smaller than the top and bottom squark masses. Since
the gluino mass should be larger than 140 GeV, then, to have bottom
mass corrections  smaller than 10$\%$, the heavy squark
mass values should be larger than 1--2 TeV
(observe that, since we are assuming universality
of the gaugino masses, the gluino mass constraint may be directly
inferred  from the chargino mass constraints, without relying on the
CDF bounds).
Therefore, the top quark mass may be close to its
fixed point value, but  a very heavy squark spectrum is needed.
Observe that this is a general statement based only  on the assumption
of  dominance of the scalar soft supersymmetry breaking
terms and has no dependence on the supersymmetry breaking
scheme. Larger gaugino masses do not help, since they not
only increase the bottom mass corrections, but they also
make $|m_{H_2}^2|$ larger.

Away from the infrared fixed point,
the situation becomes more relaxed. In general,
the closer one gets to the fixed point value, the heavier should
the spectrum be.
In order to minimize the bottom mass corrections with
low values of the squark
masses, the following condition needs to be fulfilled:
\begin{equation}
m_{H_2}^2(0) \left( \frac{7}{3} \frac{Y_f}{Y} -1 \right) \simeq
m_Q^2(0) + m_U^2(0) - \frac{7}{10} \frac{Y_f}{Y}
\left( 1 - \frac{5 Y}{7 Y_f}
- \left( 1 - \frac{Y}{Y_f} \right)^{5/7} \right) \Delta_{DU},
\label{eq:stcond}
\end{equation}
where $\Delta_{DU} = m_D^2(0) - m_U^2(0) + m_{H_1}^2(0)
-  m_{H_2}^2(0)$. The above condition ensures that
$\mu^2 \simeq m_{H_2}^2 \ll m_{\tilde{q}}^2$.
In addition, the condition
$m_{H_1}^2 > m_{H_2}^2$ implies
\begin{equation}
m_{H_1}^2(0) - m_{H_2}^2(0) > \frac{1}{6} M_{E_{eff}}^2(0)
\left[1-\left(1-\frac{Y}{Y_f}\right)^{1/4}\right] +
\frac{3}{5}
\left[1-\left(1-\frac{Y}{Y_f}\right)^{5/7}\right]  \Delta_{DU}\; .
\end{equation}
Finally, in order to avoid tachyonic solutions in the squark sector,
we need,
\begin{eqnarray}
m_U^2(0) & > & \frac{2}{3} m_{H_2}^2(0),
\nonumber\\
m_Q^2(0) & > & \frac{2}{3} m_{H_2}^2(0) + \frac{1}{5}
\Delta_{DU} \left( 1 - \left( 1 - \frac{Y}{Y_f} \right)^{5/7}
\right)
\nonumber\\
m_D^2(0) & > & \frac{2}{3} m_{H_2}^2(0) + \frac{2}{5}
\Delta_{DU} \left( 1 - \left( 1 - \frac{Y}{Y_f} \right)^{5/7}
\right).
\end{eqnarray}
%
%

The requirement
that $\mu^2$ is not correlated with the squark masses,
Eq. (\ref{eq:stcond}), implies,
for a given top quark mass,
a very precise relationship between the different supersymmetry
breaking parameters at the grand unification scale,
which may
only be fulfilled in particular supersymmetry breaking scenarios.
The necessity of  fulfilling the additional constraints given
above reduces even more
the degree of arbitrariness of the supersymmetry
breaking parameters at high energies. The further away from the
top quark mass fixed point we are, the easier the conditions are
to  fulfil. For example, for a top quark mass $M_t \simeq
175$ GeV, which corresponds to $Y/Y_f \simeq 0.8$, the above
conditions may be fulfilled by a particularly simple set
of supersymmetry breaking parameters: All the scalars acquire
a universal soft supersymmetry breaking term $m_0^2$, apart from
$m_{H_1}^2(0) \simeq 2 m_0^2$.

One may wonder about the source of the non--universality of the
soft supersymmetry breaking scalar masses.
In principle, the SO(10) symmetry assures that
all squark and sleptons belonging to a single
representation of SO(10) acquire
 a common soft supersymmetry breaking term
$m_0^2$, while the two Higgs doublets, belonging to a
10 of SO(10), have a common supersymmetry breaking
mass $m_H^2(0)$ at the grand unification scale. There
may be, however, additional sources of supersymmetry
breaking at the SO(10) breaking scale. A particular natural one
is the presence of a $D$--term associated with the necessary
$U(1)_X$ gauge symmetry breakdown to reduce the rank of the
SO(10) group to that of SU(5). If this term were present,
it would
break supersymmetry in a very specific way. The following
boundary conditions would be obtained in this case \cite{HRS2},
\begin{eqnarray}
m_{H_1}^2(0) & = & m_H^2(0) + 2 m_{X}^2
\;\;\;\;\;\;\;\;\;\;\;\;\;\;\;\;\;\;\;\;\;\;\;\;\;\;\;\;\;\;\;\;
m_{H_2}^2(0) = m_H^2(0) - 2 m_{X}^2
\nonumber\\
m_{U}^2(0) & = & m_Q^2(0) = m_E^2(0) =
m_0^2 +  m_{X}^2 \;\;\;\;\;\;\;\;\;\;\;\;
m_{D}^2(0) = m_L^2(0) = m_0^2 - 3 m_{X}^2,
\end{eqnarray}
where $m_{X}$ is the extra supersymmetry breaking
term associated with the SO(10) gauge symmetry breaking.
Observe that  scalar fields transforming with the same quantum numbers
under $SU(5)$ have degenerate masses.
In this specific case, $\Delta_{DU} = 0$ and the above
conditions take a particularly simple form. In particular,
Eq. (\ref{eq:stcond}) leads to
\begin{equation}
2 m_0^2 \simeq m_H^2(0) \left( \frac{7 Y_f}{3 Y} - 1 \right)
- \frac{14 Y_f}{3 Y} m_{X}^2.
\end{equation}
In addition, the requirement $m_{H_1}^2 > m_{H_2}^2$ puts a
lower bound on $m_{X}^2$, while the requirement of a
positive $m_D^2$ puts an upper bound on $m_{X}^2$; these
are given by
\begin{equation}
\frac{72 Y / Y_f}{7  \left[1 - \left(1 - Y/Y_f \right)^{1/4} \right]}
+ 2 > \frac{m_{H}^2(0)}{m_{X}^2} >  \frac{ 2 \left( 1
+ 5 Y/7 Y_f\right)}{(1 - Y/Y_f)}.
\end{equation}
The above conditions cannot be fulfilled for $Y/Y_f > 0.88$
($M_t \geq 185$--190 GeV for $\alpha_3(M_Z) \simeq 0.12$--0.13),
while for $Y/Y_f \simeq 0.8$ ($M_t \simeq
 175$--180 GeV) they are
only fulfilled for $m_{H}^2(0) \simeq 20 \; m_{X}^2$
($m_0^2 \simeq 0.8 m_H^2(0)$).

\section{Conclusions}

In this talk, we have analysed the radiative
breakdown of the electroweak symmetry within the Minimal
Supersymmetric Standard  Model, in the case in which  the three
Yukawa couplings of the third generation unify at the grand unification
scale.
We have shown that, due to large bottom quark mass corrections
induced through the supersymmetry breaking sector of the theory at
the one--loop level, the top quark mass predictions depend
strongly on the nature of the soft supersymmetry breaking parameters at
low energies. Moreover,
the chargino contribution to the bottom mass corrections
is strongly correlated with the contribution of these  sparticles to the
$b \rightarrow s \gamma$ decay rate.

In the  interesting case of universal soft supersymmetry
breaking parameters at the grand unification scale, the requirement
of electroweak symmetry breaking leads to strong correlations between
the mass parameter $\mu$ and the gaugino masses. The universal
scalar mass is constrained to be smaller than the universal gaugino mass,
with $M_{1/2}$ larger than 300 GeV. The result is a heavy sparticle
spectrum with strong correlations between the different mass parameters.
These correlations allow a precise computation of the bottom mass
corrections, which turn out to be very significant. In fact,
an upper bound on the top quark mass is derived,
$M_t \leq$ 165 (175) (185) GeV
for  $ \alpha_3(M_Z) \simeq$ 0.12 (0.125) (0.13). Moreover,
in order to have the
$b \rightarrow s \gamma$  decay rate
within its experimental bounds, the common
gaugino mass needs to be further constrained, $ M_{1/2} > $ 550 GeV.
This implies a very heavy squark and gaugino
spectrum, with the lightest squark mass
larger than 1 TeV. A relatively light
Higgs spectrum is obtained, with
the lightest CP-even Higgs mass in the range 110--130 GeV and a charged
Higgs mass bounded to be
$C_{\tilde{q}}  m_{\tilde{q}} \geq m_{H^{\pm}} \geq$
180 GeV, where  $C_{\tilde{q}} = 0.15$--0.2 and
$m_{\tilde{q}}$ is the characteristic third
generation squark masses. For squark masses lower than 2 TeV,
for example, the charged Higgs mass cannot be larger than 400 GeV.

The above
 conditions may be modified if non-universal soft supersymmetry
breaking parameters at the grand unification scale are considered.
We have shown that it is possible to minimize the supersymmetric bottom
mass corrections and the contributions to the
$b \rightarrow s \gamma$ decay rate, which allows to accommodate a
heavier top quark mass.
This demands the presence of light gauginos in the spectrum.
The squarks and sleptons
are in general heavy, with masses larger than
or of the order of 1 TeV.
A very heavy squark and slepton spectrum may
only be avoided if very specific relations between the soft supersymmetry
breaking parameters are fulfilled. Nevertheless,
these relations may only be
satisfied if there is a departure of  the
top quark mass  from its infrared fixed point value.
For example, if the only source of non--universality is associated
with the breakdown of  the SO(10) symmetry down to SU(5) via
the corresponding U(1) $D$--term, the necessary requirements to avoid
a heavy squark spectrum may be fulfilled only if
$M_t \leq$ 185--190 GeV for $\alpha_3(M_Z) \simeq 0.12$--0.13,
and for a very specific relation among the high energy mass
parameters, $m_0^2 \simeq 0.8 m_{H}^2(0) \simeq 16 m_{X}^2$,
where $m_0^2$ and $m_H^2(0)$ are the universal squark and
Higgs mass parameters before the SO(10) breakdown, while
$m_X^2$ is the $D$--term contribution.
In the general case  the condition of
bottom--top Yukawa coupling unification requires specific relations among
the   soft supersymmetry breaking parameters of the theory, which demand
a  departure from the infrared fixed point solution of the
top quark mass in order to allow moderate values for the squark mass
spectrum.
{}~\\
{}~\\
{}~\\
{\bf{Acknowledgements.}}  The results presented here were partially
obtained in collaboration with M. Olechowski and S. Pokorski, to whom
we are grateful.  We would also like to thank S. Dimopoulos, S. Raby,
U. Sarid and R. Rattazzi
for useful discussions. This work  is  partially
supported by the Worldlab.
{}~\\

\end{document}